\documentclass[preprint,superscriptaddress,showpacs,preprintnumbers,amsmath,amssymb]{revtex4}

\usepackage{graphicx}
\usepackage{dcolumn}
\usepackage{bm}
\usepackage{color}

\newcommand{\lb}{\left(}
\newcommand{\rb}{\right)}
\newcommand{\ls}{\left[}
\newcommand{\rs}{\right]}
\newcommand{\Lb}{\left\{}
\newcommand{\Rb}{\right\}}

\newcommand{\rket}{\right\rangle}
\newcommand{\lvl}{\left|}

\newcommand{\bet}{\beta}

\newcommand{\vep}{\varepsilon}

\newcommand{\kap}{\kappa}
\newcommand{\lam}{\lambda}

\newcommand{\sig}{\sigma}

\newcommand{\vph}{\varphi}

\newcommand{\Del}{\Delta}

\newcommand{\hatr}{\hat{\br}}

\newcommand{\balp}{\mbox{\boldmath$\alpha$}}
\newcommand{\br}{\mbox{\boldmath$r$}}
\newcommand{\bp}{\mbox{\boldmath$p$}}

\newcommand{\benu}{\begin{enumerate}}
\newcommand{\eenu}{\end{enumerate}}
\newcommand{\beq}{\begin{equation}}
\newcommand{\eeq}{\end{equation}}
\newcommand{\beqn}{\begin{eqnarray}}
\newcommand{\eeqn}{\end{eqnarray}}

\begin{document}


\title{Avoid the Tsunami of the Dirac sea in the Imaginary Time Step method}

\author{Ying Zhang}
 \affiliation{State Key Lab Nucl. Phys. {\rm\&} Tech., School of Physics, Peking University, Beijing 100871, China}

\author{Haozhao Liang}%
 \affiliation{State Key Lab Nucl. Phys. {\rm\&} Tech., School of Physics, Peking University, Beijing 100871, China}
 \affiliation{Institut de Physique Nucl\'eaire, IN2P3-CNRS and Universit\'e Paris-Sud,
    F-91406 Orsay Cedex, France}

\author{Jie Meng}
 \affiliation{State Key Lab Nucl. Phys. {\rm\&} Tech., School of Physics, Peking University, Beijing 100871, China}
 \affiliation{Department of Physics, University of Stellenbosch, Stellenbosch, South Africa}

\date{\today}

\begin{abstract}

The discrete single-particle spectra in both the Fermi and Dirac sea
have been calculated by  the imaginary time step (ITS)  method for
the Schr\"{o}dinger-like equation after avoiding the "tsunami" of
the Dirac sea, i.e., the diving behavior of the single-particle
level into the Dirac sea in the direct application of the ITS method
for the Dirac equation. It is found that by the transform from the
Dirac equation to the Schr\"{o}dinger-like equation, the
single-particle spectra, which extend from the positive to the
negative infinity, can be separately obtained by the ITS evolution
in either the Fermi sea or the Dirac sea. Identical results with
those in the conventional shooting method have been obtained via the
ITS evolution for the equivalent Schr\"{o}dinger-like equation,
which demonstrates the feasibility, practicality and reliability of
the present algorithm and dispels the doubts on the ITS method in
the relativistic system.

\end{abstract}


\maketitle


The imaginary time step (ITS) method~\cite{ev8-80}, which searches
for the direction of the steepest descent of energy and follows it
in iterative steps until the local minimum on the energy surface is
reached, is an effective approach to solve the non-relativistic
problems in coordinate space and has achieved lots of success in the
conventional mean field~\cite{ev8-cpc05}.  As noted already in
Ref.~\cite{ev8-80}, since only the lowest eigenstate is reached
after several exponential transformations, the method is applied to
the single-particle Hamiltonians of which the spectrum is bounded
from below.

From the physics point of view, a lot of novel aspects of nuclear
structure and entirely unexpected features have been found in the
"exotic nuclei"~\cite{Tanihata85,RIB01,Jonson04,Jensen04}. The
extreme neutron richness of exotic nuclei and the physics related to
the low density in the tails of matter distributions have not only
attracted attention in nuclear physics and astrophysics but also
provided a challenge to solve complex many-body problem in
coordinate space.

As one of the best candidates for the description of exotic nuclei,
the Relativistic Mean Field approach~\cite{Serot86} has achieved
lots of success in describing many nuclear phenomena during the past
years~\cite{Ring96,Vretenar05,MengPPNP}.

Subsequently, the Relativistic Continuum Hartree-Bogoliubov (RCHB)
theory~\cite{MengHaloPRL} provides a fully self-consistent treatment
of pairing correlations in the presence of the continuum and thus a
reliable description of nuclei far away from the line of
$\beta$-stability, which well reproduced the halo phenomena in
$^{11}$Li. A natural extension of the RCHB theory is the exploration
of deformed exotic nuclei and to pin down the existence of deformed
halo. Efforts along this line have been made in the past several
years. Due to the difficulty in solving the coupled channel
differential equations for deformed system in coordinate
space~\cite{Price87}, the method for expansion in Woods-Saxon basis
was proposed~\cite{ZhouWS03}, which becomes very time consuming for
the heavy system.

For the ITS method, the extension from spherical to deformed systems
is straightforward. However, whether the ITS method can be applied
for the relativistic system still remains an interesting question
due to the existence of the Dirac sea. Although, by implementing the
effective Hamiltonian deduced for the upper component to the
existing ITS routine for the Skyrme-Hartree-Fock approach, efforts
have been made to get the single-particle levels in the Fermi sea,
e.g., in Ref.~\cite{Gogelein07}, further works are still necessary
to justify its feasibility and reliability, and to understand the
influence of the Dirac sea explicitly.

In this paper, the ITS method will be applied to solve the Dirac
equation in coordinate space. The disaster of the Dirac sea
confronted in the direct application of the ITS method for the Dirac
equation will be analyzed and discussed in detail. Reliable and
practical recipe to avoid the disaster will be provided by examining
the relation between the Dirac equation and its "equivalent"
Schr\"{o}dinger-like equation.

The main idea and detailed formalism of the ITS method can be found
in Ref.~\cite{ev8-80}. The ITS evolution of the single-particle wave
functions reads
 \beq
   \lvl\Psi_a^{(n+1)}\rket
   = {\lb 1-\lam\hat{h}\rb}\lvl\vph_a^{(n)}\rket,
   \quad \lam \equiv\Delta t/\hbar,\label{ImTimeEvol}
 \eeq
where the exponential series is truncated at the first order as in
Ref.~\cite{ev8-cpc05}, $\hat{h}$ is the single-particle Hamiltonian
and \{$\vph_a^{(n)}$\} is a set of orthogonal single-particle wave
functions at time $t_n=n\Delta t$. Orthonormalizing
\{$\Psi_a^{(n+1)}$\} thus obtained generates a new set of orthogonal
wave functions \{$\vph_a^{(n+1)}$\} which will be used for the next
ITS evolution. The total energy of the system will decrease as the
time evolves, until it finds a local minimum on the energy surface.
The corresponding wave functions \{$\vph_a$\} will simultaneously
converge to the eigenstates of the single-particle Hamiltonian
$\hat{h}$.

The ITS method has been successfully implemented in the conventional
mean field approach. For the relativistic case, the main issue is to
solve the Dirac equation,
  \beq
       \Lb \balp\cdot\bp+\bet \ls M+S(\br)\rs+V(\br)\Rb \vph_a
       = (\vep_a + M ) \vph_a,
        \label{Dirac-eigen1}
  \eeq
with the scalar and vector potentials $S(\br)$ and $V(\br)$. For
clarity and simplicity, $V(\br)\pm S(\br)$ are assumed to be the
spherical Woods-Saxon potentials similar as in Ref.~\cite{ZhouWS03},
then the Dirac spinor takes the form,
 \beq
   \vph_a= \displaystyle\frac{1}{r}\left( \begin{array}{c}
                    iF_a(r)\mathcal{Y}_a(\hatr)\\
                    G_a(r)\hat{\sig}_r\mathcal{Y}_a(\hatr)
                  \end{array}
           \right)\chi_{\frac{1}{2}}(q_a),
 \eeq
with the isospinor $\chi_{\frac{1}{2}}(q_a)$, and the spherical
spinor $\mathcal{Y}_a(\hatr)$. Throughout this paper, the
single-particle state in either Fermi or Dirac sea is labeled by the
quantum numbers $\{n_a,l_a,\kap_a,j_a\}$ of the upper component,
where $\kap_a=\pm|j_a+1/2|$ for $l_a=j_a \pm 1/2$. The radial Dirac
equation for $F_a$ and $G_a$ can thus be obtained as
 \beq
  \left(\begin{array}{cc}
          V+S   &  \displaystyle-\frac{d}{dr}+\frac{\kap_a}{r} \\
         \displaystyle \frac{d}{dr}+\frac{\kap_a}{r} & V-S-2M
        \end{array}\right)
  \left(\begin{array}{c}
         F_a(r) \\
         G_a(r)
        \end{array}\right)
  =\vep_a \left(\begin{array}{c}
         F_a(r) \\
         G_a(r)
        \end{array}\right),  \label{radialDirac}
 \eeq
in which the $2\times 2$ matrix operator on the l.h.s. is the
single-particle Hamiltonian $\hat{h}$ to be used in
Eq.~(\ref{ImTimeEvol}) for the ITS evolution.

For the initial orthogonal wave functions \{$\vph_a^{(1)}$\}, the
upper components are taken as spherical Bessel function and the
lower components are assumed to be zero. The first(second)-order
differential operator in the single-particle Hamiltonian is
calculated by the seven(nine)-points formula as in
Ref.~\cite{ev8-cpc05}. The ITS evolution is carried out in
coordinate space within a spherical box $[0,R]$ with the mesh size
$dr=0.1 \sim 0.5$~fm and  a typical time step $\Del t=10^{-26}$~s.


      \begin{figure}
       \includegraphics[width=0.5\textwidth]{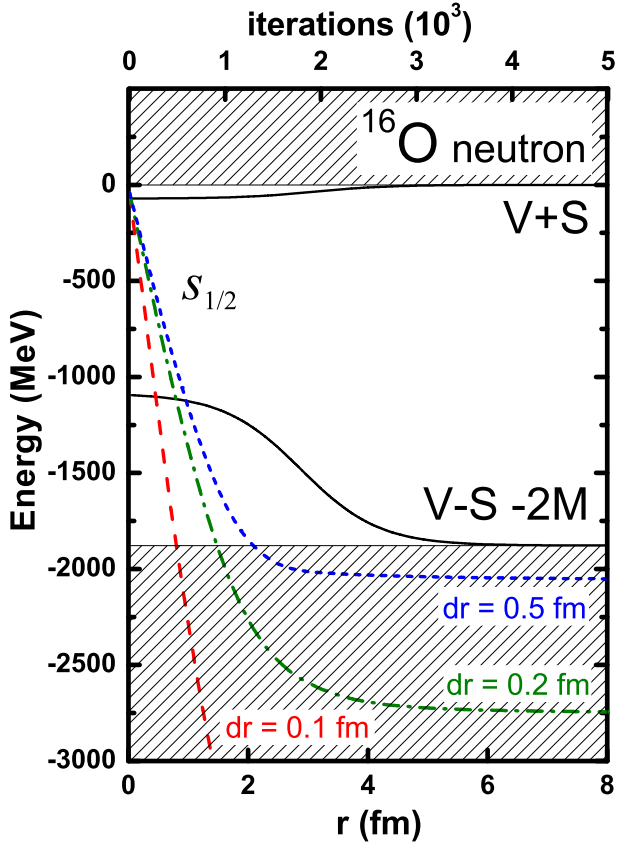}
        \caption{(color online) Disaster for solving the Dirac equation with scalar and vector Woods-Saxon potentials
                 in the ITS method. Taking $^{16}$O as an example,
                 the diving behaviors into the Dirac sea during the ITS evolution are shown
                 for the neutron single-particle level $s_{1/2}$, which is evolved
                 in a spherical box with $R=20$ fm,
                 and mesh sizes $dr=0.5$ fm (short-dashed line), $0.2$ fm (dash-dotted line),
                 and $0.1$ fm (dashed line) respectively. The shaded areas are the continuum in the Fermi and Dirac sea.
                 } \label{Fig1}
      \end{figure}

  Taking the Woods-Saxon potentials for $V(\br)\pm S(\br)$
  corresponding to $^{16}$O as an example, the ITS evolutions of the lowest $s_{1/2}$
  orbital for neutron are shown in Fig.~\ref{Fig1}.
  It is found that although the initial state is set in the Fermi sea,
  the single-particle energy will inevitably dive into the Dirac sea
  as the time evolves.
  This is not surprising. Because the ITS method searches for the states
  with the lowest energy in the whole spectra,
  the single-particle energy should dive to the negative infinity
  corresponding to the wave function with infinite nodes.
  In practice, however, as the node number is limited by the given box and
  mesh sizes, one can always achieve convergence for the ITS
  evolution. The convergent single-particle energy will
  dive deeper in the Dirac sea for smaller mesh size.

\begin{figure}
       \includegraphics[width=0.5\textwidth]{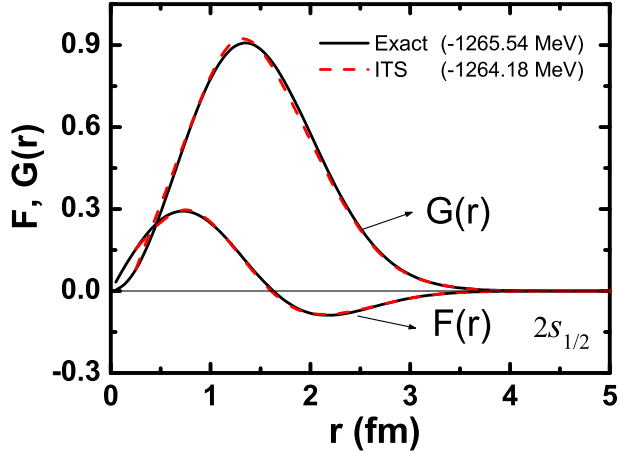}
       \caption{(color online) Neutron radial wave functions and single-particle energy
                for the survivor $2 s_{1/2}$ in the disaster of the Dirac sea.
                The ITS evolution for the Dirac equation is performed in a spherical box with $R=10$ fm and $dr=0.5$ fm,
                while the "exact" results are obtained by the shooting method with $R=20$ fm and
                $dr=0.1$ fm.
       } \label{Fig2}
      \end{figure}

  One may wonder whether the ITS evolution is able to produce
  any $s_{1/2}$ orbital outside the Dirac sea.
  Attempts have been made but a dilemma has been found.
As indicated in Fig.~\ref{Fig1}, if one takes a large mesh size, the
lowest $s_{1/2}$ orbital may lie shallow in the Dirac sea, and the
whole spectrum is sparse. Taking a box with $R = 10$ fm and $dr =
0.5$ fm as an example, a survivor $2 s_{1/2}$ is found in the
discrete region of the Dirac sea, and is compared with the "exact"
results obtained by the shooting method~\cite{MengPPNP} with $R =
20$ fm and $dr = 0.1$ fm in Fig.~\ref{Fig2}. Its single-particle
energy and radial wave functions are in good agreement with the
"exact" ones, which shows that at least some physical solutions in
the Dirac sea are accessible by the direct ITS evolution for the
Dirac equation. However, it is at a cost of precision, and it is
gloomy to go beyond the Dirac sea. To calculate more single-particle
states, one should take smaller mesh size or larger box, which
determines the number of orthogonal wave functions. However, the
whole spectrum of $s_{1/2}$ orbitals becomes denser and dives
deeper, thus it is incurable to go beyond the Dirac sea.

Therefore, this is indeed a great disaster to perform the ITS
evolution directly for the Dirac equation, since physical solutions
are completely submerged by the mighty flood with negative energy
states, or the "tsunami" of the Dirac sea.

In order to avoid the "tsunami", one possibility is to perform the
ITS evolution for the Schr\"{o}dinger-like equation instead of the
Dirac equation. Naively, such possibility might be doubted as the
Schr\"{o}dinger-like equation is deduced from the Dirac equation,
and they are equivalent in the sense that all the solutions of the
Dirac equation satisfy the corresponding Schr\"{o}dinger-like
equation. However,  they are not equivalent in the sense discussed
below.

From Eq.~(\ref{radialDirac}), the relation between the upper and
lower components can be obtained as,
 \beq
  G_a = \frac{1}{2M_+}\lb \frac{dF_a}{dr}+\frac{\kap_a}{r}F_a \rb,
   \label{FtoG}
 \eeq
with the effective mass $M_+ = M+(S-V+\vep_a)/2$, and the
corresponding Schr\"{o}dinger-like equation for the upper component
is $\hat{h} F_a = \vep_a  F_a$, with the effective Hamiltonian,


 \begin{equation}
  \hat{h} = -\frac{1}{2M_+}\frac{d^2}{dr^2}
  +\frac{1}{2M_+^2}\frac{dM_+}{dr}\frac{d}{dr}
  + \ls (V+S)+\frac{1}{2M_+^2}\frac{dM_+}{dr}\frac{\kap_a}{r}
  +\frac{1}{2M_+}\frac{\kap_a(\kap_a+1)}{r^2}
  \rs.\label{heffF}
 \end{equation}

Now the difference between the Dirac equation and the corresponding
Schr\"{o}dinger-like equation is clear. For the Dirac equation, the
single-particle spectra extend from the positive to the negative
infinity. While for the Schr\"{o}dinger-like equation, although the
effective Hamiltonian in Eq.~(\ref{heffF}) which contains the
single-particle energy is quite sophisticated, it can be reduced as
$\hat{h} = \displaystyle\frac {P^2} {2 M_+ ( \vep_a)} + U ( \vep_a)
$. Therefore, with the trial single-particle energy $\vep_a$ in the
Fermi sea, the positive definite effective mass will ensure that
$\vep_a \geq U ( \vep_a)$. As a result, the single-particle spectrum
in the Fermi sea can be separated from that in the Dirac sea, and
then the ITS evolution could be applied for the Schr\"{o}dinger-like
equation to obtain the solutions in the Fermi sea. Furthermore, in
order to obtain the solutions in the Dirac sea, one could apply the
similar technique to perform the ITS evolution for the charge
conjugated Schr\"{o}dinger-like equation. Therefore, it is the
transform from the Dirac equation to the Schr\"{o}dinger-like
equation that allows the ITS evolution for the single-particle
spectra in either the Fermi sea or the Dirac sea, and thus provides
a "life buoy" to avoid the disaster of the Dirac sea.

Using the effective Hamiltonian in Eq.~(\ref{heffF}), the upper
component is obtained by the ITS evolution for the
Schr\"{o}dinger-like equation. The corresponding lower component is
obtained by Eq.~(\ref{FtoG}).  Then the Dirac spinor including both
the upper and lower components are orthonormalized for the next
iteration.


      \begin{figure}
       \includegraphics[width=0.5\textwidth]{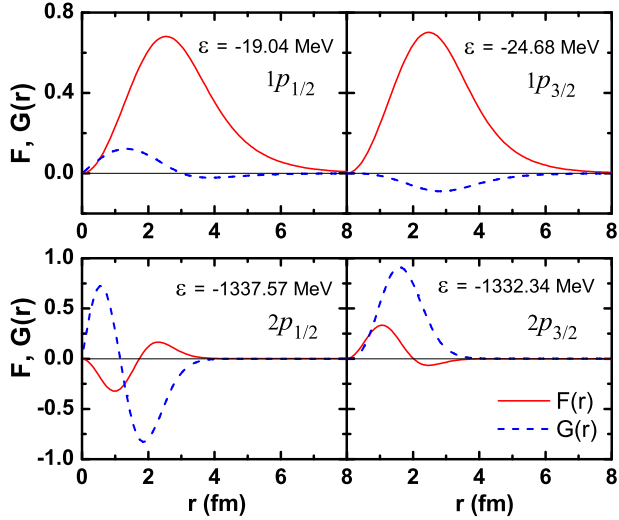}
       \caption{(color online) Neutron radial wave functions and single-particle energies
                for $1p$ spin doublets in the Fermi sea (upper panels)
                and $2p$ spin doublets in the Dirac sea (lower panels)
                obtained by the ITS evolution for the Schr\"{o}dinger-like
                equation with $R=20$ fm and $dr=0.1$ fm.
        }\label{Fig3}
      \end{figure}

As examples, the neutron radial wave functions and single-particle
energies for $1p$ spin doublets in the Fermi sea and $2p$ spin
doublets in the Dirac sea obtained by the ITS evolution for the
Schr\"{o}dinger-like equation with $R=20$ fm and $dr=0.1$ fm are
respectively given in the upper and lower panels in Fig.~\ref{Fig3}.
Both the wave functions and single-particle energies are identical
with the "exact" results obtained by the shooting method.

      \begin{figure}
       \includegraphics[width=0.5\textwidth]{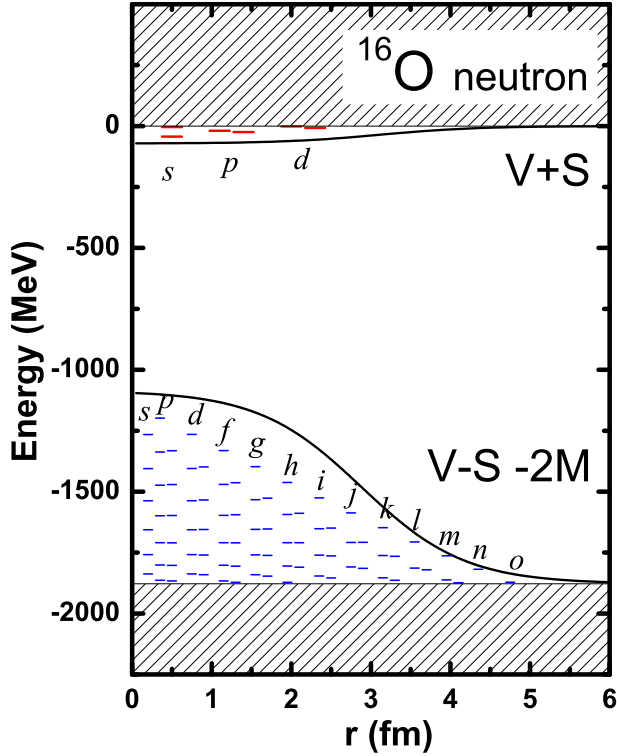}
       \caption{ (color online)
                Neutron discrete single-particle spectra of $^{16}$O in the Fermi and Dirac sea
                obtained after avoiding the disaster of the Dirac sea
                via the ITS evolution for the Schr\"{o}dinger-like equation.
                For each pair of the spin doublets, the
                left (right) is for $\kappa > 0$ ($< 0$).
       }\label{Fig4}
      \end{figure}

Similarly, all the discrete single-particle spectra identical with
the "exact" ones for neutron of $^{16}$O in both the Fermi and Dirac
sea are obtained as shown in Fig.~\ref{Fig4}. The spin doublets are
presented in pairs, with the left~(right) one is for $\kap>0~(<0)$.
It should be noted that the single-particle states are always
labeled by the quantum numbers of the upper component, different
from those in Ref.~\cite{ZhouPRL03}. Therefore, it is clearly shown
that the disaster of the Dirac sea can be avoided via the ITS
evolution for the equivalent Schr\"{o}dinger-like equation.

In conclusion, the discrete single-particle spectra in both the
Fermi and Dirac sea have been calculated by the ITS evolution for
the Schr\"{o}dinger-like equation after avoiding the "tsunami" of
the Dirac sea.

The ITS method, which has been widely used in the conventional mean
field approaches, will inevitably meet a serious challenge when it
is directly applied for the Dirac equation. All the single-particle
levels will dive into the Dirac sea during the ITS evolution, and
the physical solutions of the Dirac equation are submerged by the
"tsunami" of the Dirac sea. Fortunately, the Schr\"{o}dinger-like
equation deduced from the Dirac equation provides the life buoy to
survive the "tsunami". For a long time, the Schr\"{o}dinger-like
equation was generally regarded to be equivalent to the Dirac
equation. However, by the transform from the Dirac equation to the
Schr\"{o}dinger-like equation, the single-particle spectra, which
extend from the positive to the negative infinity, can be separately
obtained by the ITS evolution in either the Fermi sea or the Dirac
sea. The ITS evolution for the Schr\"{o}dinger-like or charge
conjugated Schr\"{o}dinger-like equation will provide the solution
in the Fermi or Dirac sea, which are identical with the "exact" ones
obtained by the shooting method.

The successful implementation here fully demonstrated the
feasibility, practicality and reliability for the ITS method in the
relativistic system.

\section*{Acknowledgements}
The authors are grateful to P. -H. Heenen, G.~C.~Hillhouse, P.~Ring,
H.~Sagawa, N.~Van~Giai, and D.~Vretenar for helpful discussion. This
work is partly supported by Major State 973 Program 2007CB815000,
the NSFC under Grant Nos. 10435010, 10775004 and 10221003.


\end{document}